\documentclass[11pt]{article}
\usepackage{amssymb,dsfont,amsmath,amsfonts,mdframed,mathrsfs,stmaryrd,bbold}

\usepackage[hyphens]{url}

\usepackage[hidelinks]{hyperref}
\hypersetup{breaklinks=true,
    colorlinks = true,
             linkcolor = blue,
            urlcolor  = blue,
            citecolor = red,
            anchorcolor = blue  
         }

\usepackage{color}
\usepackage{ulem}
\usepackage{xfrac}

\usepackage[active]{srcltx}
\usepackage{slashed}
\usepackage{color}

\usepackage{hyperref}
\usepackage{breakurl}
\hypersetup{breaklinks=true}

\advance \textheight by \topskip
\let\oldsqrt\sqrt
\def\sqrt{\mathpalette\DHLhksqrt}
\def\DHLhksqrt#1#2{%
\setbox0=\hbox{$#1\oldsqrt{#2\,}$}\dimen0=\ht0
\advance\dimen0-0.2\ht0
\setbox2=\hbox{\vrule height\ht0 depth -\dimen0}%
{\box0\lower0.4pt\box2}}

\DeclareMathAlphabet{\mathpzc}{OT1}{pzc}{m}{it}

\numberwithin{equation}{section}

\begin{document}

\markboth{Authors' Names}
{Instructions for Typing Manuscripts (Paper's Title)}
\newcommand{\nc}{\newcommand} 

\nc{\be}{\begin{equation}} 
\nc{\ee}{\end{equation}} 
\def\bsp#1\esp{\begin{split}#1\end{split}} 
\nc{\beqa}{\begin{eqnarray}} 
\nc{\eeqa}{\end{eqnarray}} 
\nc{\nn}{\nonumber} 
\nc{\noi}{\noindent} 
\def\bpm{\begin{pmatrix}} 
\def\epm{\end{pmatrix}} 
 \nc{\B}{\Big|}

\nc{\R}{{\cal R}} 
\nc{\E}{{\cal E}}
\nc{\G} {{\cal G}} 

\nc{\ie}{{\it i.e.}} 
\nc{\eg}{{\it e.g.}} 
\nc{\etc}{{\it etc.}}
\nc{\hc}{{\text{h.c.}}} 
\nc{\e}{\varepsilon}
\nc{\la}{\big<}
\nc{\ra}{\big>} 
\def\g{{\mathfrak g}} 
\nc{\D}{{\cal D}} 
\nc{\tD}{\widetilde{\D}}
\nc{\cD}{\check \D}
\nc{\Dbar}{{\overline \D}}
\nc{\Wbar}{{\overline W}}
\nc{\Dhat}{{\hat D}}
\nc{\Fhat}{{\hat F}}
\nc{\del}{{\partial}} 
\nc{\lag}{{\cal L}}
\nc{\lla}{\langle}
\nc{\rra}{\rangle}
\nc{\EE}{{\cal E}}
\nc{\F}{{ {}^* \hskip -.08truecm F}}
\renewcommand{\d}{{\rm d}} 
 
\nc{\alphadot}{{\dot\alpha}} 
\nc{\betadot}{{\dot\beta}} 
\nc{\deltadot}{{\dot\delta}} 
\nc{\edot}{{\dot\epsilon}} 
\nc{\gammadot}{{\dot\gamma}} 
\nc{\etadot}{{\dot\eta}} 
\nc{\thetabar}{{\bar\theta}} 
\nc{\Thetabar}{{\bar\Theta}} 
\nc{\Sigmabar}{{\Sigma^\dag}} 
\nc{\psibar}{{\bar\psi}} 
\nc{\etabar}{{\bar\eta}} 
\nc{\ebar}{{\bar\e}} 
\nc{\xibar}{{\bar\xi}} 
\nc{\chibar}{{\bar\chi}} 
\nc{\lambar}{{\bar\lambda}} 
\nc{\sibar}{{\bar\sigma}}

\nc{\is}{{i^\ast}} 
\nc{\js}{{j^\ast}} 
\nc{\ks}{{k^\ast}} 
\nc{\ms}{{m^\ast}} 
\nc{\ls}{{\ell^\ast}} 
 
\nc{\tm}{{\tilde \mu}} 
\nc{\tn}{{\tilde \nu}} 
\nc{\tr}{{\tilde \rho}} 
\nc{\ts}{{\tilde \sigma}} 
\nc{\ta}{{\tilde \alpha}} 
\nc{\tb}{{\tilde \beta}} 
\nc{\taa}{{\widetilde{\dot \alpha}}} 
\nc{\tbb}{{\widetilde{\dot \beta}}} 
\nc{\tM}{{\tilde M}} 
\nc{\bM}{{\bar M}} 
\nc{\tN}{{\tilde N}} 
\nc{\tP}{{\tilde P}} 
\nc{\tQ}{{\tilde Q}}

\nc{\rc}{\textcolor{red}}
\nc{\bc}{\textcolor{blue}}
\nc{\ma}{\textcolor{magenta}}


\title{A pedagogical discussion of $N=1$ four-dimensional supergravity in superspace
}

\author{
Robin Ducrocq$^1$\footnote{robin.ducrocq@iphc.cnrs.fr }, Michel Rausch de Traubenberg$^1$\footnote{michel.rausch@iphc.cnrs.fr}, Mauricio Valenzuela$^2$\footnote{supercuerda@gmail.com}\\[12pt]
$^1$ {\scriptsize IPHC-DRS, UdS, CNRS, IN2P3}\\[-2pt]
{\small 23  rue du Loess, Strasbourg, 67037 Cedex, France}\\[5pt]
$^2$ {\small Centro de Estudios Cient\'{\i}ficos (CECs)}\\[-2pt]
{\small Arturo Prat 514, Valdivia, Chile } 
}


\maketitle


\begin{abstract}
A short introduction to $N=1$ supergravity  in four dimensions in the superspace approach is given emphasising on all steps
to obtain the final Lagrangian.
In particular starting from geometrical principles and the introduction of superfields in curved superspace, the  action coupling matter and gauge fields to supergravity  is derived.
This  review is
based  on the book {\rm ``A supergravity Primer: From geometrical principles to the Final Lagrangian"} \cite{mm} and on several lectures given at the doctoral
school of Strasbourg.
\vspace{5pt}

\noindent \texttt{keywords: {Supergravity, Superfield, Superspace}}

\end{abstract}


\section{Introduction}

 The purpose of this brief review is to provide the conceptual scheme that  leads to the four-dimensional $N=1$ supergravity Lagrangian,   from  geometrical principles and the definition of a {\it had hoc curved} superspace. The technical details will be omitted
since we would like to emphasise on basic ideas. 

Among the   many different approaches we shall follow the one of Wess and Bagger \cite{wb}. An alternative approach can be found in, \eg \ , the  book \cite{fp} in which  superconformal techniques are used.

 There are many references on supergravity, among which we would like to mention  the first paper on the subject devoted to pure supergravity \cite{fnf} and the subsequent developments which include the coupling of matter fields \cite{csj1}. Due to limited number of pages of this review we will not be able to quote all the important articles that have been key to  the understanding  of supergravity. For a complete list of references the reader may consult  \eg\  the books \cite{wb,mm,fp}.

This review closely follows the book \cite{mm} of two of us where the reader will be able to deepen in the technical aspects,  given there in great detail. In order to facilitate the reading the title of each section corresponds to the same title of the corresponding chapter of the book. The book \cite{mm} can also be  supplemented with the book \cite{rf} (devoted to a pedagogical
introduction $-$ in French $-$ of supersymmetry), these two books resulted from lectures and seminars given during several years at the doctoral school of Strasbourg.

\section{Curved superspace}

The essential idea to consider a superspace in supergravity is  to add fermionic directions enhance  the usual bosonic spacetime coordinates.  The curved bosonic and fermionic  coordinates are covariant under  general { coordinate} transformations.  However, in the flat tangent space  the bosonic (resp. fermionic) coordinates are the standard  vector (resp. left and right handed spinors),
transforming accordingly under local Lorentz transformations, thus parametrising  the usual superspace of supersymmetry. Then, analogously as in general relativity, two elementary
superfields are needed: the supervierbein and the spin-connection, together with their corresponding super-curvature and super-torsion tensors (see Section \ref{sec:vc}).
It is then shown that imposing some constraints, the so called \textit{torsion constraints}, it is possible to reduce the number of elementary superfields and then express all the
torsion and curvature tensors in terms of  three fundamental superfields (see Section \ref{sec:b}). Fixing the gauge as done in Section \ref{sec:gf} enables us  to identify the gravity multiplet.
The lowest component of the superfields are given in Section \ref{sec:lc}. The superconformal group or Weyl supergroup is also identified in Section \ref{sec:W}.

\subsection{Supervierbein, superconnection, curvature and torsion}
\label{sec:vc}
The first step is to define two different frames: the Einstein frame and the local tangent
frame or Lorentz frame (see \ref{app:conv} for conventions). 
In the local tangent frame, a point is parametrised by
 $z^M=(x^\mu, \theta^\alpha, \thetabar_\alphadot)$ and we have covariance under local Lorentz transformations:
\beqa
x'^\mu = x^\nu \Lambda_\nu{}^\mu(z) \ , \ \ 
\theta'^\alpha=\theta^\beta \Lambda_\beta{}^\alpha(z) \ ,\ \ 
\thetabar'_\alphadot =\thetabar_\betadot \Lambda^\betadot{}_\alphadot(z) \ . 
\label{eq:lorentz}
\eeqa
The matrices $ \Lambda_M{}^ N (z)= ( \Lambda_\mu{}^\nu(z), \Lambda_\alpha{}^\beta(z), \Lambda^\betadot{}_\alphadot(z))$ are superfields,
corresponding to the vector, left-handed and right-handed spinor representations respectively,  depend on the point $z$.
Note that transformations in the tangent space do not mix
indices of different nature. 
 Furthermore, the Lorentz generators associated to the corresponding representation obey the relations,
\beqa
\label{eq:Lor}
J_{\mu \nu}&=& \frac 14 (\sibar_\nu)^{\alphadot \alpha} (\sibar_\mu)^{\betadot \beta}(\e_{\alpha\beta} J_{\alphadot\betadot}-\e_{\alphadot\betadot} J_{\alpha\beta})\ , \\
J_\alpha{}^\beta&=& (\sigma^{\nu \mu})_\alpha{}^\beta J_{\mu\nu}\ , \nn\\
J^\alphadot{}_\betadot&=& (\sibar^{\nu \mu})^\alphadot{}_\betadot J_{\mu\nu}\ . \nn
\eeqa
We also introduce the $z^M$-conjugated  variables 
$\partial_M =(\partial_\mu, \partial_\alpha, \bar \partial^\alphadot)$ which together satisfy  the graded commutation relations
\beqa
[\partial_M, z^N]_{|M||N|} \equiv \partial_M z^N - (-1)^{|M||N|}
z^N \partial_M = \delta_M {}^N \ ,
\eeqa
where $|\mu|=0, |\alpha|=1, |\alphadot|=1 $ is the grading
of the index $M =(\mu,\alpha,\alphadot)$.

In the Einstein (or \textit{curved} space) frame, a  point is parametrised by $z^\tM=(x^\tm,\theta^\ta, \thetabar_\taa)$ and we have invariance under  superdiffeomorphisms
\beqa
\label{eq:sudef}
z'^\tM=z^\tM + \xi^\tM(z) \ ,
\eeqa
where $\xi^\tM(z)$ are superfields depending on the point $z$.

The first dynamical superfield is the supervierbein 
$E_\tM{}^N(z)$ and its inverse $E_M{}^\tN(z)$ which connect components in the Einstein  frame to
the components in the Lorentz (or flat) frame. Thus \eg\ for a vector superfield:
\beqa
V_\tM = E_\tM{}^N V_N\ , \quad 
V_M = E_M{}^\tM V_\tM \ . \nn
\eeqa
Considering the metric in the flat tangent space $\eta_{MN}=(\eta_{\mu\nu}, \epsilon_{\alpha \beta}, \epsilon^{\alphadot\betadot})$ we thus have
  \beqa
g_{\tM \tN} = (-)^{|\tM| ( |N| + |\tN|)} E_\tN{}^N E_\tM{}^M \eta_{MN} \ . \nn 
  \eeqa
\medskip
The second dynamical variable is the superconnection 
$\Omega_{\tM MN}$ with symmetry property
\beqa
\Omega_{\tM MN}= -(-)^{|M| |N|} \Omega_{\tM NM} \ , \nn
\eeqa
which is in addition Lie algebra valued, \ie, is non-vanishing only when $M$ and $N$ are of the same nature. The superconnection  has the transformation law
\beqa
\Omega'_{\tM M}{}^{N} = (\Lambda^{-1})_M{}^P \Omega_{\tM P}{}^Q \Lambda_Q{}^ N    - (\Lambda^{-1})_M{}^P \partial_\tM 
\Lambda_P{}^N \ ,\nn
\eeqa
under a local Lorentz transformation and it enables to define the
covariant derivatives
\beqa \label{eq:cov}\begin{array}{l l} 
  \D_\tM X^M = \del_\tM X^M + (-)^{|\tM| |N|} X^N \Omega_{\tM N}{}^M\ , & 
  \quad \D_\tM X_M = \del_\tM X_M -  \Omega_{\tM M}{}^N X_N\ ,\\[5pt]
  \D_N X^M = E_N{}^\tM \D_\tM X^M \ , & 
  \quad \D_N X_M = E_N{}^\tM \D_\tM X_M \ .
\end{array} \eeqa

The closure of the algebra leads to
\beqa
\big[\D_M, \D_N\big]_{|M||N|} = T_{MN}{}^P \D_P -\frac12
R_{MN \underline{\alpha} \underline{\beta}} J^{\underline{\beta} \underline{\alpha}} \ , \nn
\eeqa
where $\underline{\alpha} = (\alpha, \alphadot)$,  and
 we only sum over independent generators
say $J_{\alpha \beta}$ and $J^{\alphadot \betadot}$ (see equation \eqref{eq:Lor}).
The torsion and curvature fulfill the obvious symmetry properties
\beqa \label{eq:sym} 
  T_{MNP} &=&-(-)^{|M||N|} T_{NMP}\ , \nn\\
  R_{MNPQ}&=& -(-)^{|M||N|} R_{NMPQ} \ ,\\
    R_{MNPQ}&=& -(-)^{|P||Q|} R_{MNQP} \ . \nn
    \eeqa
    Introducing
\beqa
 T_{\tP \tQ}{}^Q &=& 
    (-)^{|\tP|(|N|+|\tQ|)} E_\tQ{}^N E_\tP{}^M T_{MN}{}^Q\ , \nn\\
     R_{\tM \tN S}{}^Q &=& 
    (-)^{|\tM|(|N|+|\tN|)} E_\tN{}^N E_\tM{}^M R_{MNS}{}^Q
    \ , \nn
\eeqa  
    a little algebra leads to
\beqa 
\label{eq:TR}
  T_{\tP \tQ}{}^Q &=& -\D_\tP E_\tQ{}^Q +(-)^{| \tP | |\tQ |} \D_\tQ E_\tP{}^Q \
     ,  \\ 
  R_{\tM \tN S}{}^Q &=&  
    \del_\tM \Omega_{\tN S}{}^Q 
     - (-)^{|\tM| |\tN|}\del_\tN \Omega_{\tM S}{}^Q   
     - \Omega_{\tM S}{}^R \Omega_{\tN R}{}^Q 
     + (-)^{|\tM||\tN|} \Omega_{\tN S}{}^R \Omega_{\tM R}{}^Q  \nn \ . 
     \eeqa

\subsection{The Bianchi identities}\label{sec:b} 
The covariant derivatives satisfy the Bianchi identities          
\beqa \label{eq:bian} 
  0= &&
    (-)^{|M_1||M_3|} \Big[ \D_{M_1}, \big[\D_{M_2},\D_{M_3} \big]_{|M_2| 
     |M_3|}\Big]_{|M_1|(|M_2|+|M_3|)}  \nn\\ 
  && (-)^{|M_2||M_1|} \Big[\D_{M_2},\big[\D_{M_3},\D_{M_1}\big]_{|M_3| 
     |M_1|}\Big]_{|M_2|(|M_3|+|M_1|)}  \\ 
 && (-)^{|M_3||M_2|} \Big[\D_{M_3},\big[\D_{M_1},\D_{M_2}\big]_{|M_1| 
     |M_2|}\Big]_{|M_3|(|M_1|+|M_2|)} \ . \nn
\eeqa 
Developing the double graded commutators  
using the Lie algebra valuedness of the curvature tensor and the property
$
  [X,Y Z]_{|X|(|Y| +|Z|)} = [X,Y]_{|X||Y|} Z + (-)^{|X||Y|} Y[X,Z]_{|X||Z|} \ , \nn
$
leads to two
series of identities:
\beqa \label{eq:id} 
    0=&&
    (-)^{|M_1||M_3|} \Big[\D_{M_1} T_{M_2 M_3 S} - T_{M_1 M_2}{}^R T_{R M_3 S} 
      + R_{M_1 M_2 M_3 S}\Big] \nn\\ 
    &&   
      (-)^{|M_2||M_1|} \big[\D_{M_2} T_{M_3 M_1 S} - T_{M_2 M_3}{}^R T_{R M_1 S} 
      + R_{M_2 M_3 M_1 S}\Big] \nn\\\ 
    &&
      (-)^{|M_3||M_2|} \big[\D_{M_3} T_{M_1 M_2 S} - T_{M_3 M_1}{}^R T_{R M_2 S} +
       R_{M_3 M_1 M_2 S}\Big] \ , \nn\\
    0=&& 
    (-)^{|M_1||M_3|} \Big[T_{M_1 M_2}{}^R R_{R M_3 P Q} -  
      \D_{M_1} R_{M_2 M_3 P Q} \Big] \nn\\\ 
     && 
    (-)^{|M_2||M_1|} \Big[T_{M_2 M_3}{}^R R_{R M_1 P Q} -  
      \D_{M_2} R_{M_3 M_1 P Q} \Big] \nn\\ 
     &&  
    (-)^{|M_3||M_2|} \Big[T_{M_3 M_1}{}^R R_{R M_2 P Q} - 
      \D_{M_3} R_{M_1 M_2 P Q} \Big] \ .  \
    \eeqa
    These  Bianchi identities are usual.     
    However, since this geometrical construction leads to a large number of superfields,
    additional
   constraints are necessary to reduce the number of physical degrees of freedom. These constraints are also necessary 
   to be able to construct minimal supergravity models, \ie,  which realise the supersymmetry algebra in flat limit. It turns out that one possible set of constraints
   is the following:
   {
     \beqa \label{eq:cT}
     \hskip -.2truecm
  T_{\underline{\alpha} \underline{\beta}}{}^{\underline{\gamma}}=
T_{\alpha \beta}{}^\mu = T_{\alphadot \betadot}{}^\mu=
  T_{\underline{\alpha} \mu}{}^\nu = T_{\mu \underline{\alpha}}{}^\nu=
    T_{\mu \nu}{}^\rho=0  ,
    & T_{\alpha \alphadot}{}^\mu = T_{\alphadot \alpha}{}^\mu = -2i 
    \sigma^\mu{}_{\alpha \alphadot}\ .
\eeqa
   }
  
In particular the constraints \eqref{eq:cT} are compatible with the definition of chiral superfields (Section \ref{sec:field}).
These series of constraints allow to  reduce the Bianchi identities and express   all torsion and curvature
tensors in terms of fewer independent quantities. Thus we end up with three independent superfields, namely  the chiral surperfield 
${\cal R}$,  the real vector superfield $G_\mu$ and  the symmetric chiral superfield $W_{(\alpha \beta \gamma)}$.
To obtain the explicit expressions of all torsion and curvature tensors, we need to perform a lengthy computation that we do not reproduce here (for further details see reference \cite{mm}), but we  point out  key observations to get the final results: 
\begin{enumerate}
    \item  The second series of identities in \eqref{eq:id} are consequences of
the first series of identities.
\item Accounting of all symmetries the first series of identities leads to thirteen different equations
(in fact thirty but only     thirteen are independent). 
 \item The main idea  in order to  reduce the Bianchi identities  is to convert  all vector indices into spinor indices.  For instance  for a component of the curvature tensor
\beqa
  \sigma^\mu{}_{\gamma\gammadot} \sigma^\nu{}_{\delta\deltadot} R_{\alphadot
    \betadot \mu \nu} =  
  R_{\alphadot \betadot \gamma \gammadot \delta \deltadot} =  
  2 \e_{\gamma \delta} R_{\alphadot \betadot \gammadot \deltadot} - 
  2 \e_{\gammadot \deltadot} R_{\alphadot \betadot \gamma \delta} \ , \nn
\eeqa 
where the second equality is a consequence of the Lie algebra valuedness of the curvature tensor. Similarly for a component of the  torsion tensor
 we get
\beqa \label{eq:irT} 
  \sigma^\mu{}_{\gamma\gammadot}  T_{\alphadot \mu \beta}=  T_{\alphadot \gamma \gammadot \beta}=  
        \e_{\alphadot\gammadot} \e_{\gamma \beta} T
      + \e_{\gamma \beta} T_{(\alphadot \gammadot)}  
      + \e_{\alphadot \gammadot} T_{(\gamma \beta)}  
      + T_{(\alphadot \gammadot)(\gamma \beta)} \ , 
\eeqa
where the new $T$-tensors are symmetric with respect to the exchange of the
indices in the parentheses and correspond to the decomposition of $T_{\alphadot \gamma \gammadot \beta}$ into irreducible representations of $\mathfrak{sl}(2,\mathbb C) \cong \mathfrak{so}(1,3)$.
\end{enumerate}

As stated previously all non-vanishing torsion and curvature tensors can be expressed in terms of the superfields ${\cal R}, G_\mu$ and $W_{(\alpha \beta \gamma)}$. The non-vanishing brackets are given by 
\beqa
\label{eq:comTR}
  \Big\{\D_\alpha, \D_\beta\Big\} &=&-  
    \frac12 R_{\alpha\beta\gamma\delta} J^{\delta\gamma}\ , \nn\\
  \Big\{\overline \D_\alphadot, \overline \D_\betadot\Big\} &=& -  
    \frac12 R_{\alphadot\betadot\gammadot\deltadot} J^{\deltadot\gammadot} \ , \nn\\
  \Big\{\D_\alpha,\overline \D_\alphadot\Big\} &=& T_{\alpha\alphadot}{}^\mu \D_\mu -
    \frac12 R_{\alpha\alphadot\beta\gamma} J^{\gamma\beta}-
    \frac12 R_{\alpha\alphadot\betadot\gammadot} J^{\gammadot\betadot}\ , \nn\\
  \Big[\D_\alpha,\D_\mu\Big] &=&T_{\alpha\mu}{}^\beta \D_\beta -
    T_{\alpha\mu}{}^\betadot\ \overline \D_\betadot -
    \frac12 R_{\alpha\mu\beta\gamma} J^{\gamma\beta}-
    \frac12 R_{\alpha\mu\betadot\gammadot} J^{\gammadot\betadot} \ , \\
  \Big[\overline \D_\alphadot,\D_\mu\Big] &=& T_{\alphadot\mu}{}^\beta \D_\beta -
    T_{\alphadot\mu}{}^\betadot\ \overline \D_\betadot -
    \frac12 R_{\alphadot\mu\beta\gamma} J^{\gamma\beta}-
    \frac12 R_{\alphadot\mu\betadot\gammadot} J^{\gammadot\betadot} \ , \nn\\
  \Big[\D_\mu,\D_\nu\Big] &=& T_{\mu\nu}{}^\beta \D_\beta -
    T_{\mu\nu}{}^\betadot\ \overline \D_\betadot -
    \frac12 R_{\mu\nu\beta\gamma} J^{\gamma\beta}-
    \frac12 R_{\mu\nu\betadot\gammadot} J^{\gammadot\betadot}\ ,\nn
    \eeqa
(see \eqref{eq:cov} for the relationship between derivatives with Einstein and Lorentz indices).
We do not give all the expression of the $R-$ and $T-$ tensors. The 
reader may refer to Tables 3.2, 3.3 and 3.4 of \ \cite{mm} for explicit expressions and constraints upon the various tensors.

\subsection{The Weyl supergroup}\label{sec:W}
The torsion constraints \eqref{eq:cT} and the corresponding algebra \eqref{eq:comTR} have   additional symmetries. Indeed,  the automorphism group of the algebra, or equivalently the set of  transformations
upon the generators $\D_M$ and $J_{MN}$ which leave \eqref{eq:comTR}  invariant is the super-Weyl  or superconformal group. This is the analogue of the Weyl group of general relativity corresponding to Weyl rescaling of the metric. We now determine the transformations $\delta J_{\underline{\alpha} \underline{\beta}}$ and $\delta \D_M$ which leave \eqref{eq:comTR} invariant. The commutation relations of the Lorentz algebra leads obviously to $\delta J_{\underline{\alpha} \underline{\beta}}=0$. Setting
\beqa
\label{eq:Lort}
\delta \D_\alpha= - \Phi \D_\alpha +\D^\beta \Phi J_{\alpha \beta} 
\eeqa
and similar expressions for $\Dbar_\alphadot, \D_\mu$ we obtain after lengthy algebraic manipulations 
\beqa \label{eq:ht}
  \delta_\Sigma \D_\alpha &=& \big[ \Sigma - 2  \Sigma^\dag\big] \D_\alpha 
    -\D^\gamma \Sigma J_{\alpha \gamma} \ , \nn\\
  \delta_\Sigma \Dbar_\alphadot &=& \big[ \Sigma^\dag -2  \Sigma\big] \Dbar_\alphadot 
     +\Dbar^\gammadot \Sigma^\dag J_{\alphadot \gammadot} \ , \\
  \delta_\Sigma \D_\mu &=& -(\Sigma +  \Sigma^\dag) \D_\mu 
     -\frac{i}2 \big[ \Dbar \Sigma^\dag \bar\sigma_\mu\D + 
      \D\Sigma \sigma_\mu \Dbar\big] - \D^\nu(\Sigma +  \Sigma^\dag) J_{\mu \nu} \ ,\nn\\
  \delta_\Sigma J_{{\alpha} {\beta}} &=&\ 0  \ , \nn\\
  \delta_\Sigma J_{\alphadot\betadot}&=&0\ , \nn
\eeqa
where $\Sigma$ is a chiral superfield, \ie,  satisfying
$\Dbar_\alphadot \Sigma=0$.

Using the explicit expressions (not given here) of the torsion and curvature tensors together with the algebra 
\eqref{eq:comTR} we obtain
\beqa
\label{eq:Sig-RGW}
 \delta_\Sigma \R &=& - 4 \Sigma\R -\frac14 (\Dbar\!\cdot\!\Dbar - 8
  \R) \Sigmabar \ , \nn\\
   \delta_\Sigma G_\mu &=& - \big[\Sigma + \Sigmabar\big]
    G_\mu + i \D_\mu\big[\Sigma-\Sigmabar\big] \ ,\\
     \delta_\Sigma W_{(\alpha\beta\gamma)} &=& - 3 \Sigma W_{(\alpha\beta\gamma)} \ .\nn
\eeqa
It can be seen that the superfield $\delta_\Sigma {\cal R}$ is chiral as it should (see Section
\ref{sec:chiral}).

Finally from the definition of  covariant derivatives
\eqref{eq:cov} we obtain
\beqa
\label{eq:tOm}
  \delta_\Sigma \Omega_{\tM \beta\gamma} &=&
   E_{\tM \beta} \D_\gamma \Sigma + 
    E_{\tM \gamma} \D_\beta \Sigma - E_{\tM
    \nu}\e_{\gamma\alpha} (\sigma^{\nu
    \rho})_\beta{}^\alpha \D_\rho \big[\Sigma+\Sigma^\dag\big] \ , \nn \\
  \delta_\Sigma \Omega_{\tM \betadot\gammadot}& =&\
      E_{\tM \betadot} \Dbar_\gammadot \Sigmabar 
     + E_{\tM \gammadot} \Dbar_\betadot \Sigmabar 
     - E_{\tM \nu} \e_{\betadot\alphadot}
       (\sibar^{\nu\rho})^\alphadot{}_\gammadot
       \D_\rho \big[\Sigma+\Sigma^\dag\big] \ ,\nn \\
  \delta_\Sigma \Omega_{\tM \nu \rho}& =& 
     -E_{\tM \nu} \D_\rho \big[\Sigma +  \Sigma^\dag\big]
    + E_{\tM \rho} \D_\nu \big[\Sigma+\Sigma^\dag\big] \nn\\
&&    - 2 E_\tM{}^\alpha (\sigma_{\nu\rho})_\alpha{}^\beta
      \D_\beta \Sigma
    - 2 E_{\tM\alphadot}
      (\sibar_{\nu\rho})^\alphadot{}_\betadot \Dbar^\betadot \Sigmabar \ ,
\eeqa
and
\beqa \label{eq:SigmaE}
  \delta_\Sigma E_{\tM}{}^\alpha &=& \big[2 \Sigma^\dag -\Sigma\big] E_\tM{}^\alpha 
   + \frac{i}{2} E_\tM{}^\mu \big(\Dbar \Sigma^\dag \sibar_\mu)^\alpha\ , \nn\\ 
  \delta_\Sigma E_{\tM \alphadot} &= & \big[2\Sigma -\Sigma^\dag\big] E_{\tM
   \alphadot} + \frac{i}2 E_\tM{}^\mu (\D\Sigma \sigma_\mu)_\alphadot \ ,  \\
  \delta_\Sigma E_{\tM}{}^\mu &= & (\Sigma+\Sigma^\dag) E_\tM{}^\mu  \  .\nn
\eeqa

\subsection{Supergravity transformations and gauge fixing conditions}
\label{sec:gf}
We have expressed all quantities in terms of a few number
of superfields. Nonetheless using the large symmetry due to the  supergravity algebra many components can be  set to
zero  by means of an appropriated choice of parameters of the  symmetry transformations \eqref{eq:lorentz} and \eqref{eq:sudef}. Under a Lorentz transformation \eqref{eq:lorentz} and a superdiffeomorphism  \eqref{eq:sudef} we have
\beqa
  V^M(z) &\ \ \to\ \ &  V'{}^M(z') = V^N(z) \Lambda_N{}^M(z)\ , \nn \\
  E_\tM{}^M(z) &\to&  E'{}_\tM{}^M(z') =
    \frac{\del z^\tN}{\del z'{}^\tM} E_\tN{}^N(z) \Lambda_N{}^M(z) \ , \\
  \Omega_{\tM P}{}^Q(z)&\to & \Omega'{}_{\tM P}{}^Q(z') =
    \frac{\del z^\tN}{\del z'{}^\tM}
    (\Lambda^{-1})_P{}^R(z) \bigg[ \Omega_{\tN R}{}^S(z) \Lambda_S{}^Q(z) -
    \del_\tN \Lambda_R{}^Q(z) \bigg] \ . \nn
\label{eq:GRLtrans}\eeqa
At the infinitesimal level with $\Lambda^M{}_N = \delta^M{}_N + L^M{}_N$ and with the following redefinition 
\beqa
 L_P{}^Q \to L_P{}^Q -\xi^{\tM} \Omega_{\tM P}{}^Q\ ,\nn
\eeqa
we obtain
\beqa \label{eq:sugra} 
  \delta V^M &=& -\xi^N \D_N V^M + V^N
    L_N{}^M \ , \nn\\  
  \delta E_\tM{}^M &=& -\D_\tM \xi^M + \xi^\tN T_{\tN \tM}{}^M + E_\tM{}^P
    L_P{}^M \ ,  \\   
  \delta \Omega_{\tM P}{}^Q &=& -\xi^\tN R_{\tN \tM P}{}^Q + \Omega_{\tM P}{}^R
    L_R{}^Q - L_P{}^R \Omega_{\tM R}{}^Q -\del_\tM L_P{}^Q \ .\nn
\eeqa
Similarly we have for  the superfields
${\cal R}$ and $G^\mu$ 
\beqa \label{eq:RG} 
  \delta {\cal R} = - \xi^\tM \D_\tM {\cal R}  \ , \ \ 
  \delta G^\mu = - \xi^\tM \D_\tM G^\mu + G^\nu L_\nu{}^\mu \ .\nn
\eeqa

In the computation below we will see that  only the lowest component in the Taylor expansion of the various superfields will be relevant.
From now on, given a superfield $X$ we set
$X\big| \equiv X\big|_{\theta=\thetabar=0}$ for the lowest component of $X$ in its Grassmann expansion.
It can be easily seen, using the huge arbitrariness of the parameters, that   some of the lowest components can be  fixed to zero. 
In particular  for the superconnection
\beqa \label{eq:Ome0} 
  \Omega_{\tm M}{}^N\Big| = \omega_{\tm M}{}^N (x)\ ,  \ \ 
  \Omega_{\ta M}{}^N\Big|=0  \ , \ \
  \Omega^{\taa}{}_M{}^N\Big| = 0 \ ,
\eeqa
and for the supervierbein and its inverse
\beqa \label{eq:E0} 
  E_\tM{}^M(z)\Big| &= &\bpm 
    e_\tm{}^\mu(x) & \frac12 \psi_\tm{}^\alpha(x) & \frac12
      \psibar_{\tm\alphadot}(x) \\ 
    0 & \delta_\ta{}^\alpha & 0 \\ 
    0 & 0 & \delta^\taa{}_\alphadot
  \epm \ , \nn\\ 
  E_M{}^{\tM}(z)\Big| &=& \bpm 
    e_\mu{}^\tm(x) & -\frac12 \psi_\mu{}^\ta(x) & -\frac12 \psibar_{\mu\taa}(x) \\ 
    0 &\delta_\alpha{}^\ta & 0 \\ 
    0 & 0 & \delta^\alphadot{}_\taa
   \epm  \ , 
   \eeqa
   where $e_\tm{}^\mu$ is the helicity-two  graviton, \ie, the vierbein,  whereas the Majorana spinor-vector $(\psi_\tm{}^\alpha,
      \psibar_{\tm\alphadot})$ is the gravitino. Using
      \eqref{eq:E0} it is important for the sequel to observe the relationships
     \beqa\label{eq:redsugramult}
 & e_\tm{}^\mu e_\mu{}^\tn = \delta_\tm{}^\tn\ , \qquad  
  e_\mu{}^\tm e_\tm{}^\nu = \delta_\mu{}^\nu\ ,\nn\\
&  \psi_\mu{}^\ta = e_\mu{}^\tm \psi_\tm{}^\alpha \delta_\alpha{}^\ta\ , \qquad
  \psibar_{\mu \taa} = e_\mu{}^\tm \psibar_{\tm\alphadot}
    \delta^\alphadot{}_\taa\ .
\eeqa 
Finally the lowest order components of ${\cal R}$ and $G_\mu$ cannot be eliminated, thus we define
\beqa \label{eq:RG0} 
  {\cal R}(z)\Big| = -\frac16 M(x) \ , \ \ 
  G_{\mu}(z)\Big|&=& -\frac13 b_{\mu}(x) \ .
\eeqa
Following  \ref{app:conv} we define also  $b_{\alpha \alphadot} = \sigma^\mu{}_{\alpha \alphadot} b_\mu$.

Having gauge fixed the superconnection and the supervierbein as above the dynamical fields of supergravity are given by the graviton $e_\tm{}^\mu$,
the gravitino $(\psi_\tm{}^\alpha, \psibar_{\tm\alphadot})$ and two auxiliary  fields, the complex  scalar $M$  and the real vector $b_\mu$. We will see later  that the connection $\omega_{\tm M}{}^N$ is a composite field and it can be expressed in terms of the graviton and
the gravitino.

Of course the  gauge   fixing  conditions \eqref{eq:Ome0} and
\eqref{eq:E0} are not preserved by  \eqref{eq:sudef} and \eqref{eq:lorentz}. We thus have to restrict the set of transformations. We assume at first
\beqa \label{eq:para0} 
  \xi^\mu(z)\B          = 0\ , \qquad 
  \xi^\alpha(z)\B       = \e^\alpha(x)\ , \qquad
  \xibar_\alphadot(z)\B = \ebar_\alphadot(x)  \ , \qquad
  L_{MN}(z)\B           = 0 \ .
\eeqa 
Then in order to preserve the gauge fixing condition, a supergravity transformation reduces to the following  combination of  superdiffeomorphism and local Lorentz transformation
\beqa
\label{eq:Tsugra}
 && \xi^\mu(z) =2i \Big[\tilde \theta\sigma^\mu\ebar - \e \sigma^\mu\tilde\thetabar 
    \Big]\ ,\qquad 
  \xi^\alpha(z) = \e^\alpha\ , \qquad
  \xibar_\alphadot(z) = \ebar_\alphadot \nn \ ,\\
 && L_{\alpha \beta}(z) = \frac13\Big[ \tilde \theta_\alpha\big(2 \e_\beta
    M^\ast + b_{\beta \gammadot} \ebar^\gammadot \big) +
    \tilde \theta_\beta\big(2 \e_\alpha M^\ast + b_{\alpha\gammadot}
    \ebar^\gammadot\big) \Big]\ ,  \\
 && L_{\alphadot\betadot}(z) = \frac13\Big[ \tilde \thetabar_\alphadot\big(2 \ebar_\betadot
    M + \e^\gamma b_{\gamma\betadot}\big) + \tilde \thetabar_\betadot\big(2
    \ebar_\alphadot M + \e^\gamma b_{\gamma\alphadot}\big) \Big] \ ,\nn\\
 && L_{\mu \nu}(z) = -\frac12\Big[ (\sibar_\mu \sigma_\nu)^\alphadot{}_\betadot
    L_\alphadot{}^\betadot + (\sigma_\mu \sibar_\nu)_\alpha{}^\beta
    L^\alpha{}_\beta \Big] \nn \ ,
\eeqa 
where the fields $(\e,\ebar)$, $b$ and $M$ depend only on the spacetime
coordinates $x$ and all indices are flat superspace indices. We have also introduced
$\tilde \theta_\alpha= \theta^{\tilde \alpha} E_{\tilde \alpha}{}^\beta \e_{\alpha \beta} \ne \theta_\alpha$ as
well as $\tilde\thetabar^\alphadot$. 

\subsection{Lowest order component fields}\label{sec:lc}
In the previous subsection we have fixed the lowest component of the dynamical fields of supergravity, say the supervierbein and the superconnection. 
 It is
also possible to obtain explicit expressions of the lowest component as well as the lowest component of the covariant derivatives of the superfield $\R, G_\mu$ and $W_{(\alpha\beta\gamma)}$.
The key observation to get these expressions relies on several facts.
First of all the gauge fixing condition \eqref{eq:E0} and \eqref{eq:Ome0} give
 using  \eqref{eq:TR}
 \beqa \label{eq:E-TR} 
  T_{\tm \tn}{}^\mu\Big| &= & -\del_\tm e_\tn{}^\mu + \del_\tn e_\tm{}^\mu -
    \omega_{\tm \tn}{}^\mu + \omega_{\tn \tm}{}^\mu  \ , \nn\\ 
  T_{\tm \tn}{}^\alpha\Big| &=&-\frac12 \Big( \del_\tm \psi_\tn{}^\alpha -
    \del_\tn \psi_\tm{}^\alpha - \psi_\tm{}^\beta \omega_{\tn \beta}{}^\alpha +
    \psi_\tn{}^\beta \omega_{\tm \beta}{}^\alpha \Big)
    = -\frac12 \psi_{\tm \tn}{}^\alpha \ , \\ 
  T_{\tm \tn \alphadot}\Big| &=&-\frac12 \Big( \del_\tm \psibar_{\tn\alphadot} -
    \del_\tn \psibar_{\tm\alphadot} - \psibar_{\tm\betadot}
    \omega_\tn{}^\betadot{}_\alphadot + \psibar_{\tn\betadot}
    \omega_\tm{}^\betadot{}_\alphadot \Big)
    = -\frac12 \psibar_{\tm \tn \alphadot} \ , \nn \\ 
  R_{\tm \tn \mu}{}^\nu\B  &=& \del_\tm \omega_{\tn\mu}{}^\nu - \del_\tn
    \omega_{\tm \mu}{}^\nu + \omega_{\tn \mu}{}^\rho \omega_{\tm \rho}{}^\nu -
    \omega_{\tm \mu}{}^\rho \omega_{\tn \rho}{}^\nu \ , \nn
 \eeqa 
 where
 \beqa\label{eq:redu}
&&  \omega_{\tm \tn}{}^\mu =e_\tn{}^\nu \omega_{\tm \nu}{}^\mu\ , \ \  
  \omega_{\tm \tn \tr} =e_\tn{}^\nu e_\tr{}^\rho\omega_{\tm \nu \rho}\ , \nn\\  
 && \D_\tm \psi_\tn{}^\alpha =\del_\tm \psi_\tn{}^\alpha +
    \psi_\tn{}^\beta \omega_{\tm \beta}{}^\alpha \ ,  \ \ 
%
%
  \psi_{\tm \tn}{}^\alpha = \D_\tm \psi_\tn{}^\alpha -
    \D_\tn\psi_\tm{}^\alpha\ . 
%
\eeqa
We also define,
\beqa
 \psi_{\mu \nu} = e_\mu{}^\tm e_\nu{}^\tn \psi_{\tm \tn} \qquad (\neq\D_\mu \psi_\nu - \D_\nu\psi_\mu )\ .\nn
 \eeqa
Similar expressions hold for the right-handed counterpart of the gravitino.

The next step is to express curvature and torsion tensors with Einstein indices from  
curvature and torsion tensors with Lorentz indices and to compare the result with
\eqref{eq:E-TR}. For instance we obtain in this way
\beqa
  T_{\tm \tn}{}^\mu\B &=& E_\tn{}^N E_\tm{}^M T_{MN}{}^\mu\B 
  =-\frac{i}{2} \Big(\psi_\tm\sigma^\mu\psibar_\tn - \psi_\tn \sigma^\mu
    \psibar_\tm \Big) \ ,\nn
\eeqa
which gives, comparing with the first equation \eqref{eq:E-TR},
\beqa
  \omega_{\tm \tn \tr} &=&
     - \frac12 e_{\mu \tr} \big(\del_\tm e_\tn{}^\mu - \del_\tn e_\tm{}^\mu\big) 
    + \frac12 e_{\mu \tm} \big(\del_\tn e_\tr{}^\mu - \del_\tr e_\tn{}^\mu\big) 
    - \frac12 e_{\mu \tn} \Big(\del_\tr e_\tm{}^\mu - \del_\tm e_\tr{}^\mu\Big)  
  \nn\\
  &&    + \frac{i}{4} e_{\mu \tr} \ \big(\psi_\tm\sigma^\mu\psibar_\tn-\psi_\tn\sigma^\mu\psibar_\tm\big) 
    - \frac{i}{4} e_{\mu \tm}
       \big(\psi_\tn\sigma^\mu\psibar_\tr-\psi_\tr\sigma^\mu\psibar_\tn\big)\nn\\
&&     + \frac{i}{4}e_{\mu \tn}
       \big(\psi_\tr\sigma^\mu\psibar_\tm-\psi_\tm\sigma^\mu\psibar_\tr\big) \ . \nn
\eeqa
Of course $\omega_\alpha{}^\beta=\sfrac 1 4\;\omega_{\mu\nu} (\sigma^\mu \bar \sigma^\nu)_\alpha{}^\beta$ and
$\omega^\alphadot{}_\betadot=  \sfrac 14 \;\omega_{\mu\nu} (\bar \sigma^\mu  \sigma^\nu)^\alphadot{}_\betadot$.

Passing to the superfields $\R$ and $G_\mu$, we already know $\R\big|$ and $G_\mu\big|$ (see \eqref{eq:RG0}).
Similarly, we can  obtain $W_{(\alpha\beta \gamma)}\big|$ and
the lowest component of the covariant derivatives of the three  superfields, for which  one needs the explicit
expressions of the torsion and curvature tensors  (given in \cite{mm}) appearing in \eqref{eq:comTR}.
For instance we get
\beqa
 \D_\delta G^\mu \B =
    \frac13 \big(\sigma^\nu\psibar_\nu{}^\mu\big)_\delta
    - \frac{i}{12} \e^{\mu\nu\rho\sigma} \big(\sigma_\sigma \psibar_{\nu\rho}
        \big)_\delta 
    + \frac{i}{6} \psi^\mu{}_\delta M^\ast
    + \frac{i}{6} \big(\sigma^\nu\psibar_\nu\big)_\delta b^\mu\ . \nn
\eeqa
We do not give $W_{(\alpha \beta \gamma)}\big|$ since it is not relevant here.
For the lowest component of covariant derivatives of $\R$ we refer to \eqref{eq:Rl}
where $\R$ is expanded using the new $\Theta-$variables introduced in Section
\ref{sec:GP}.
Note that the computation of $\D_\alpha \R$ is not difficult since it is related directly to $\D_\alpha G_{\mu}\B$ whilst it is tedious
to obtain $\D \cdot \D \R \B$.\\

 With  this information we can deduce the supergravity transformations of
the supergravity multiplet $e_\tm{}^\mu, (\psi_\mu{}^\alpha, \overline{\psi}_{\mu \alphadot}), M$ and $b_\mu$:

\beqa \label{eq:tSUGRA} 
 \delta e_\tm{}^\mu &=& 
  \delta E_\tm{}^\mu\B = 
  -\D_\tm \xi^\mu\B + \xi^M T_{M \tm}{}^\mu\B + E_\tm{}^\nu L_\nu{}^\mu\B 
  =-i  \Big[ \e(x) \sigma^\mu \psibar_\tm - \psi_\tm
    \sigma^\mu \ebar(x)\Big] \ , \nn\\
     \delta \psi_\tm{}^\alpha &=& 2 \delta E_{\tm}{}^\alpha\B =
  -2 \D_\tm \xi^\alpha\B + 2 \xi^M T_{M \tm}{}^\alpha\B  + 2 E_\tm{}^\beta
  L_\beta{}^\alpha \B
  \nn\\
  &=& -2 \D_\tm \e^\alpha(x) + \frac{i}{3} e_\tm{}^\nu
  \big(\bar \e(x) \sibar_\nu\big)^\alpha M + \frac{i}{3} e_\tm{}^\nu \Big[ (\e(x)
    \sigma_\rho \sibar_\nu)^\alpha b^\rho - 3 \e^\alpha(x)\ b_\nu \Big]
    \ , \nn\\
  \delta M &=&
      -6 \delta {\cal R}\B 
    =  6 \xi^M \D_M {\cal R}\B
    =  -2 \big(\e \sigma^{\mu \nu} \psi_{\mu \nu}\big) 
      + i \big(\e \sigma^\mu \psibar_\mu\big) M 
      - i \big(\e \!\cdot\! \psi_\mu\big) b^\mu \ , \nn\\
  \delta b^\mu &=&
      -3 \delta G^\mu\B 
    =  3 \xi^M \D_M G^\mu \B - 3 G^\nu L_\nu{}^\mu\B 
    = 3 \xi^\alpha \D_\alpha G^\mu \B + 3 \xibar_\alphadot \Dbar^\alphadot
         G^\mu\B  \\
  &  = &\e \sigma^\nu \psibar_\nu{}^\mu - \bar \e \sibar^\nu\psi_\nu{}^\mu
      - \frac{i}{4} \e^{\mu\nu\rho\sigma} \Big[ \e\sigma_\sigma\psibar_{\nu\rho} +
        \bar\e \sibar_\sigma\psi_{\nu\rho} \Big] 
     + \frac{i}{2} \Big[
      \e\!\cdot\!\psi^\mu M^\ast - \bar\e\!\cdot\!\psibar^\mu M \Big] \nn\\
&&     + \frac{i}{2}
      \Big[ \e \sigma^\nu\psibar_\nu + \bar\e\sibar^\nu\psi_\nu \Big] b^\mu
    + \frac14 \e^{\mu\nu\rho\sigma} \Big[ \e \sigma_\sigma \psibar_\nu -
      \bar\e\sibar_\sigma\psi_\nu\Big] b_\rho \ . \nn
\eeqa
We observe that, as a gauge field, the transformation of the gravitino
involves the covariant derivative \eqref{eq:redu}.

\section{Superfield in curved superspace}\label{sec:field}
In this section we introduce superfields in curved superspace, namely chiral and vector superfields. 
It is important to emphasise that the
torsion constraints \eqref{eq:cT} turn out to be essential to extend
the superfields of supersymmetry in the supergravity context.
Superfields depend on $z^\tM=(x^\tm, \theta^\ta, \thetabar_\taa)$, but we have to compute lowest order component
of derivatives with Lorentz indices. More precisely, 
for both  chiral or vector superfields, in order to obtain the full supergravity action,  we have to compute several lowest order components of derivatives up to the fourth order. This  computation turns out to use intensively the algebra \eqref{eq:comTR},  the lowest components of the corresponding curvature
or torsion tensors and the torsion constraints \eqref{eq:cT}. The second important feature to be used is the conversion of flat indices into curved indices as will be seen below.

\subsection{Chiral superfields}\label{sec:chiral}

Following the conventional definition in  supersymmetry, a chiral superfield $\Phi$  has three independent components: a complex scalar $\varphi$, a left-handed spinor $\chi$ and an auxiliary field $F$ defined by 
\beqa
\label{eq:Phi}
\varphi=\Phi\B \ , \ \ 
\chi_\alpha=\frac 1 {\sqrt 2} \D_\alpha \Phi\B\ , \ \ 
F= -\frac14 \D \cdot \D \Phi \B \ .
\eeqa
Similar expressions hold for the anti-chiral superfield. Observe again that these definitions involve flat index derivatives \eqref{eq:cov}
whereas the superfield depends on curved quantities. We shall come back to this point in Section \ref{sec:GP}.

The components of $\Phi$ given
  in terms of its first orders derivatives   \eqref{eq:Phi} is the starting point to   compute the higher order derivatives. 
 It is not our purpose to address all order derivatives, which is somehow cumbersome, but we would like to stress  on the key steps in these computations. In particular we should obtain, using the conversion of flat indices to curved indices:
\beqa
\D_\mu \Phi\B= E_\mu{}^\tm \B (\D_\tm  \Phi)\B + E_\mu{}^\ta\B (\D_\ta \Phi)\B + E_\mu{}_\taa\B (\D^\taa \Phi)\B \ , \nn
\eeqa
and using \eqref{eq:Phi}, and  \eqref{eq:E0}, \eqref{eq:Ome0} obtain
\beqa
\D_\mu \Phi\B= e_\mu{}^\tm\big(\partial_\tm - \frac{\sqrt 2} 2 \psi_\tm \cdot \chi\big) \varphi  \equiv \hat D_\mu \varphi\ .
\eeqa
In a similar way we deduce
\beqa
\D_\mu\D_\alpha\Phi\B 
 =\sqrt{2} e_\mu{}^\tm \bigg[ 
       \del_\tm \chi_\alpha 
     - \omega_{\tm \alpha}{}^\beta \chi_\beta 
     + \frac{1}{\sqrt{2}}  \psi_{\tm \alpha} F  
     - \frac{i}{\sqrt{2}} \big(\sigma^\nu \psibar_\tm\big)_\alpha \Dhat_\nu \phi
     \bigg]  \equiv \sqrt{2} \Dhat_\mu \chi_\alpha 
     .  \nn
\eeqa
The two derivatives $\hat D_\mu \varphi$ and $\hat D_\mu \chi$ are covariant
with respect to supergravity transformations.

Another fundamental property that we would like to  highlight is the following. Given  an anti-chiral superfield $\Phi^\dag$ then 
\beqa
\Dbar_\alphadot \big(\Dbar\cdot \Dbar - 8 \R\big) \Phi^\dag = 0 \ , \nn
\eeqa
or equivalently  the superfield 
\beqa
\label{eq:Xi}
 \Xi= \big(\Dbar\cdot \Dbar - 8 \R\big) \Phi^\dag\ ,
\eeqa 
is chiral. 
This property is central to obtain a compact chiral expression of the supergravity action (see Section \ref{sec:GP}). 
For the lowest order components of covariant derivatives of $\Xi$ we refer to \eqref{eq:Xil}
where $\Xi$ is expanded using the $\Theta-$variables introduced in Section
\ref{sec:GP}.

Finally considering a supergravity transformation \eqref{eq:Tsugra} we obtain the infinitesimal variation of a chiral superfield
\beqa
\
\label{eq:Tphi}
  \delta \Phi &=& -\xi^M \D_M \Phi = -\xi^\alpha \D_\alpha \Phi - \xi^\mu \D_\mu \Phi
  \eeqa
  or in components
  \beqa
  \label{eq:TPhi}
  \delta \phi &=& \delta \Phi \B = -\e^\alpha \D_\alpha \Phi\B = -\sqrt{2} \e
\!\cdot\! \chi \ , \nn\\
 \delta \chi_\alpha &= & \frac{1}{\sqrt{2}} \delta\big(\D_\alpha\Phi\big)\B
    = \sqrt{2}\ \Big[ \e_\alpha  F - i \big(\sigma^\mu\bar\e\big)_\alpha
      \Dhat_\mu\phi \Big] \ , \\
   \delta F &= & \frac14 \delta\big(\D\!\cdot\!\D \Phi\big) \B = 
      \frac{\sqrt{2}}{3} \e\!\cdot\!\chi M^\ast  - i \sqrt{2}
      \Dhat_\mu\chi\sigma^\mu\bar\e - \frac{\sqrt{2}}{6} b_\mu \chi
      \sigma^\mu\bar\e\  . \nn   
  \eeqa
  We observe that differently from in supersymmetry the auxiliary field does not transform as a total derivative. 
\subsection{Vector superfields}\label{sec:VF}
Given a Lie algebra $\g$ and a representation ${\frak R}$ (see  \ref{app:conv}) a vector superfield is specified by $V= V^a T_a$ and satisfies
the reality condition
\beqa
V^\dag = V \ .\nn
\eeqa
The transformation of a vector superfield under a gauge transformation is
analogous as in supersymmetry and is given by
\beqa 
\label{eq:gauge}
e^{2g V} \to e^{-2ig\Lambda} e^{2g V} e^{2i g \Lambda}  \ , 
\eeqa
where $g$ is the coupling constant and $\Lambda = \Lambda^a T_a$ with
$\Lambda^a$ chiral superfields. Further as in supersymmetry the transformation
\eqref{eq:gauge} enables to select the so-called Wess-Zumino gauge. In this
gauge many components of $V$ vanish 
\beqa
\label{eq:curv-WZ}
  V\B=0\ , \quad 
  \D_\alpha V \B = 0\ , \quad 
  \Dbar_\alphadot V \B =0\ , \quad 
  \D\!\cdot\!\D V\B = 0 \ , \quad
  \Dbar\!\cdot\!\Dbar V\B =0 \ . 
\eeqa
We define now, in a supergravity context, 
 the superfield strength tensors
\beqa\label{eq:WbarW2}
  W_\alpha = -\frac14 \Big[\Dbar\!\cdot\!\Dbar - 8 \R \Big] e^{2gV}
    \D_\alpha e^{-2gV}  
  \ , \ \ 
  \Wbar_\alphadot &= &\frac14 \Big[\D\!\cdot\!\D - 8 \R^\dag \Big] e^{-2gV}
    \Dbar_\alphadot e^{2gV} 
    \eeqa
    which is similar to its supersymmetric definition with the substitution $D\cdot D$
    ($D$ being the usual covariant derivative in supersymmetry) by $\Dbar\cdot\Dbar -8 \R$.
The operator $\Dbar\!\cdot\!\Dbar - 8 \R$   ensures that $W_\alpha$ is chiral, \ie, $\Dbar_\alphadot W_\alpha=0$ (see \eqref{eq:Xi}).
The physical degrees of freedom of the vector superfield are given by
\beqa
\label{eq:Vcomp}
&v_\mu = \frac14 \sibar_\mu{}^{\alphadot \alpha} \big[\D_\alpha,\Dbar_\alphadot\big] V\B \ , \ \ 
\lambda_\alpha = \frac{i}{2 g} W_\alpha \B \ ,  \ \ 
  \lambar_\alphadot  = -\frac{i}{2 g} \Wbar_\alphadot \B \ , \nn\\
& D = \frac{1}{4 g} \D^\alpha W_\alpha \B =  \frac{1}{4 g} \Dbar_\alphadot
    \Wbar^\alphadot\B\ ,
\eeqa
where $v_\mu$ is a real vector, $(\lambda, \lambar)$
 is a Majorana spinor and $D$ is a real auxiliary field.
 Under the gauge transformation \eqref{eq:gauge} we have
 \beqa
 \label{eq:TW}
 W_\alpha \to e^{-2ig \Lambda} W_\alpha e^{2ig \Lambda} \ . \eeqa
 
 It is not our purpose to give all derivatives of the vector superfield, but to
 point out some that will be relevant in the sequel. In particular from
 \eqref{eq:Vcomp} we obtain
 \beqa
 \D_\alpha \Dbar_\alphadot V \B = -\Dbar_\alphadot \D_\alpha V\B = v_{\alpha \alphadot} \ , \nn
 \eeqa
 with the conventions of \ref{app:conv} to convert the vector index to spinor
 indices. We then obtain
 \beqa
  \D_\mu \D_\alpha \Dbar_\alphadot V \B 
   &=& E_\mu{}^\tm \D_\tm \D_\alpha \Dbar_\alphadot V\B + 
    E_\mu{}^\ta \D_\ta \D_\alpha \Dbar_\alphadot V\B + 
    E_{\mu\taa} \Dbar^\taa \D_\alpha \Dbar_\alphadot V\B\nn \\
 & =& 
    e_\mu{}^\tm \D_\tm v_{\alpha \alphadot} + 
    i \Big[\psi_{\mu\alpha}\lambar_\alphadot +
      \psibar_{\mu\alphadot}\lambda_\alpha \Big] -
      \frac{i}{2} \sigma^\nu{}_{\alpha \alphadot} \big(\psi_\nu \sigma^\rho
      \psibar_\mu\big) v_\rho
       \equiv \ \Dhat_\mu v_{\alpha \alphadot} \ ,\nn
      \eeqa
      and
      \beqa
  \Dhat_\mu\lambda_\alpha &= &\ \frac{i}{2g} \Big( 
    \D_\mu W_\alpha + \frac{i}{2} g \sibar_\mu{}^{\betadot \beta}
    \big[\D_\beta \Dbar_\betadot V, W_\alpha \big] \Big)\B \ , \\
    &= & e_\mu{}^\tm \Big(\D_\tm \lambda_\alpha + i g \big[v_\tm,\lambda_\alpha\big]
      \Big)
  - \frac12 \big(\sigma^{\nu\rho}\psi_\mu \big)_\alpha 
      \big(\Fhat_{\nu\rho}\big) 
  + \frac{i}{2} D \psi_{\mu\alpha} \ ,
\nn\eeqa
with
$
\Fhat_{\mu\nu}  =
   \Dhat_\mu v_\nu - \Dhat_\nu v_\mu + i g \big[ v_\mu,v_\nu\big] \ ,\nn
$
or
\beqa \label{eq:hatF2}
\hat F_{\tm \tn} = e_{\tm}{}^\mu e_{\tn}{}^\nu \Fhat_{\mu\nu}  
&=& \del_\tm v_\tn - \partial_\tm v_\tn +ig [v_\tm, v_\tn]
   -\frac{i}2
  \Big[ \bar\lambda \bar \sigma_\tn \psi_\tm + \lambda \sigma_\tn \bar\psi_\tm -
 \bar\lambda \bar \sigma_\tm \psi_\tn - \lambda \sigma_\tm \bar\psi_\tn\Big ]\nn\\
 &=&F_{\tm \tn} -\frac{i}2
  \Big[ \bar\lambda \bar \sigma_\tn \psi_\tm + \lambda \sigma_\tn \bar\psi_\tm -
 \bar\lambda \bar \sigma_\tm \psi_\tn - \lambda \sigma_\tm \bar\psi_\tn\Big ]
  \ . 
  \eeqa

  The quantities we have introduced are covariant with respect to the gauge transformations and to supergravity.

For the lowest order  components of $W_\alpha$ we refer to \eqref{eq:Wl}
where $W_\alpha$ is expanded using the $\Theta-$variables introduced in Section
\ref{sec:GP}.
 \subsection{Gauge interactions of chiral superfields}\label{sec:GAUGE}
 Let $\g$ be a compact real Lie algebra that can be simple, semisimple or even reductive, \ie, can have semisimple and $\mathfrak{u}(1)$ factors,  and let
 ${\frak R}$  be a representation of $\g$ not necessarily  irreducible.
 We define the vector superfield $V$ as in Section \ref{sec:VF}
 and introduce a chiral superfield in the representation ${\frak R}$. The anti-chiral superfield is of course in the representation $\overline{\frak R}$ (the complex conjugate representation) with generators $T_a \to -T_a ^\ast$.
 Since under a gauge transformation
$
\Phi \to e^{-2i g \Lambda} \Phi \ , \nn
$
  as in usual supersymmetry
$
\Phi^\dag e^{-2ig V} \Phi \ , \nn
$
is gauge invariant. In turn this means that the Lagrangian will
involve terms of the form $\Phi^\dag e^{-2g V}$ or $e^{-2gV} \Phi$. Consequently  we  have to compute 
\beqa 
\label{eq:coup}
{\cal X} =\big(\Dbar \cdot \Dbar - 8 \R\big)\Phi^\dag e^{-2g V} \ ,  \ \
{\cal X} ^\dag =\big(\Dbar \cdot \Dbar - 8 \R) e^{-2g V} \Phi \ ,
\eeqa
which are respectively chiral and anti-chiral superfields.
For the first term, since in the Wess-Zumino gauge $V^3=0$ 
we get
\beqa 
 {\cal X} = \Xi
        - 2 g \big(\Dbar\!\cdot\!\Dbar - 8 \R\big) \Phi^\dag V
        + 2 g^2 \big(\Dbar\!\cdot\!\Dbar - 8 \R\big) \Phi^\dag V^2\ . 
\eeqa 
For the lowest order components of ${\cal X}$ we refer to \eqref{eq:Xl}
where $\cal X$ is expanded using the $\Theta-$variables introduced in Section
\ref{sec:GP}.\\

Under a supergravity transformation  \eqref{eq:Tsugra} $V$
transforms as 
\beqa
\delta V= -\xi^M \D_M V \ ,\nn
\eeqa
but does not remain in the Wess-Zumino gauge. This means that 
 \eqref{eq:Tsugra} 
 has to be combined with a gauge transformation. We do not
 give here the precise form of the transformation of the chiral and vector superfields under the combined gauge-supergravity 
 transformations. The interested reader  is refereed to \cite{mm}, Chapter 3. We just mention that all transformations
 involve now derivatives, covariant with respect to gauge symmetry.
 
\section{General principles  to construct invariant actions} \label{sec:GP}
In this section we introduce a new set of variables in order to facilitate the expansion of chiral superfields.  We then introduce the
last ingredient in order to
construct invariant actions, given in the superspace language together with their transformation under a super-Weyl rescaling.  
 K\"ahler transformations are also discussed.

\subsection{Introduction of hybrid variables}
The superfields in curved superspace introduced in Section \ref{sec:field} depend on $z^\tM=(x^\tm, \theta^\ta, \thetabar_\taa)$ 
but their components are defined through flat covariant derivatives. Consequently the field expansion is very complicated. To overcome this difficulty we introduce an alternative set of  hybrid   variables $z^{\bar M} = (x^\tm, \Theta^\alpha, \Thetabar_{\alphadot})$ where the new $\Theta-$variables carry Lorentz indices whereas $x^\tm$ carry Einstein indices.  
The new $\Theta-$variables are chosen in such a way that the expansion of a chiral superfield is defined by its covariant
derivatives:
\beqa\label{eq:PhiTh}
  \Phi(x, \Theta) &=& \Phi\B + \sqrt 2 \Theta \cdot (\D \Phi)\B -\frac14
  \Theta\cdot \Theta (\D\cdot \D \Phi)\B\nn\\
  &=&
  \phi + \sqrt{2} \Theta \!\cdot\! \chi - \Theta \!\cdot\! \Theta F \ .
\eeqa
It is now possible to rewrite the transformation \eqref{eq:Tphi}
using the $\Theta-$variables
\beqa
\label{eq:TPhiN}
\delta \Phi= -\eta^{\bar M} \partial_{\bar M} \Phi \ , 
\eeqa
where
\beqa 
\eta^{\tm} &=& 2i \Theta \sigma^{\tm} \bar \e + 
  \Theta \!\cdot\! \Theta\ 
     \bar \psi_{\tn} \bar \sigma^{\tm} \sigma^{\tn} \bar \e\ , \nn\\
  \eta^{\alpha} &=&  \e^\alpha - i  \Theta \sigma^{\tm} \bar \e
  \psi_{\tm}{}^\alpha \nn\\
 && \hskip .5 truecm
  + \Theta \!\cdot\! \Theta \Big[
      \frac13 M^\ast \e^\alpha - i  \omega_{\tm}{}^{\alpha \beta}
      (\sigma^{\tm} \bar \e)_\beta + \frac16 b_\mu  (\bar \e \bar
      \sigma^\mu)^\alpha - \frac12 \psi_{\tn}{}^\alpha \bar \psi_{\tm}
      \bar \sigma^{\tn} \sigma^{\tm} \bar \e \Big] \ .\nn
\eeqa

As we have seen in \eqref{eq:TPhi}, the higher order component of a chiral superfield
does not transform as a total derivative. In order to define invariant actions we introduce a capacity $\Delta$ with transformation law
\beqa
\label{eq:TDelta}
  \delta \Delta  = - (-)^{|\bM|} \del_{\bM}(\eta^\bM \Delta) \ ,
\eeqa 
such that
\beqa
\label{eq:EPhi}
  \delta \big(\Delta \Phi\big) 
  =- (-)^{|\bM|} \del_\bM \big(\eta^\bM \Delta
  \Phi\big) \ ,
\eeqa 
by \eqref{eq:TPhiN}. This means that the product of a chiral superfield
with a capacity is naturally invariant under a supergravity transformation.
We choose now a specific capacity  denoted $\EE$ such that $\EE\big|=e$ with
$e = \det (e_\tm{}^\mu)$. Using 
\eqref{eq:TDelta} with the transformation property of the supergravity multiplet (see \eqref{eq:tSUGRA}) we deduce step-by-step
the components of $\EE$:
\be
  \EE = e\ \Big[
    1 + i\Theta \sigma^\tm \psibar_\tm
    - \Theta\!\cdot\!\Theta \big(M^\ast + \psibar_\tm \sibar^{\tm \tn}
      \psibar_\tn \big) \Big]\ ,\nn
\ee 
with $\sigma^\tm = e_\mu{}^\tm \sigma^\mu, \sigma^{\tm \tn} = e_\mu{}^\tm e_\nu{}^\tn \sigma^{\mu \nu}.$

We recap now the $\Theta-$expansion of the basic fields needed to derive the
supergravity action.
The anti-chiral superfield corresponding 
to \eqref{eq:PhiTh} leads to  the chiral superfields
\beqa
\label{eq:Xil}
&&  \Xi= \Xi\B + \sqrt 2 \Theta \cdot (\D \Xi)\B -\frac14 \Theta \!\cdot\! \Theta  (\D \cdot\D \Xi)\B = 4 F^\dag +\frac43 M\phi^\dag 
 + \Theta \!\cdot\! \Bigg\{- 4 i \sqrt{2} \big(\sigma^\mu \Dhat_\mu\chibar\big)
\nn \\
    &&
       + \frac{2 \sqrt{2}}{3} b_\mu \big(\sigma^\mu \chibar\big)
       + \phi^\dag\Big(\frac83 \big(\sigma^{\mu \nu}\psi_{\mu \nu}\big)  
       - \frac{4 i}{3} \big(\sigma^\mu\psibar_\mu\big) M 
       + \frac{4 i}{3} \psi^\mu b_\mu\Big)\Bigg\}\\
    && - \Theta \!\cdot\! \Theta \Bigg\{
       - 4 e_\mu{}^\tm \D_\tm \Dhat^\mu \phi^\dag
       + \frac{8 i}{3} b^\mu \Dhat_\mu \phi^\dag 
       + 2 \sqrt{2} \psibar_\mu \!\cdot\! \Dhat^\mu \chibar
       - \frac{2 \sqrt{2}}3 \bar \chi \bar \sigma^{\mu \nu} \bar \psi_{\mu \nu}
       - \frac{8}{3} M^\ast F^\dag\nn \\
  && \quad
    +  \sqrt{2} i \chibar \!\cdot\! \psibar_\mu b^\mu 
    + \frac{2\sqrt{2}i}{3}\chibar \sibar^{\nu\mu}\psibar_\mu b_\nu 
    + \phi^\dag \Big(
       -\frac23 e_\mu{}^\tm e_\nu{}^\tn R_{\tm \tn}{}^{\mu \nu}\big|
       - \frac{8}9 MM^\ast 
       + \frac{4}9 b_\mu b^\mu\nn \\
&&\quad
       +\frac{4i}{3} e_\mu{}^\tm \D_\tm b^\mu
       + \frac23 \psibar_\mu \!\cdot\! \psibar^\mu M 
       + \frac23 \psi_\nu \sigma^\nu \psibar_\mu b^\mu 
       + \frac{4 i}{3} \psibar^\mu \sibar^\nu \psi_{\mu \nu}\nn\\
       && \quad
      + \frac{1}{6} \e^{\mu \nu \rho \sigma} (\psi_\mu \sigma_\sigma
     \psibar_{\nu\rho} + \psibar_\mu \sibar_\sigma \psi_{\nu\rho})\Big)
    \Bigg\} \ , \nn
  \eeqa
and
\beqa\label{eq:Xl}
{\cal X} &=& 
{\cal X} \B + \sqrt 2 \Theta \cdot (\D {\cal X})\B -\frac14 \Theta \!\cdot\! \Theta (\D \cdot\D {\cal X}) \B \\
&=&
\Xi + \Theta\!\cdot\! \Big[
  - 4 \sqrt{2} g \big(\sigma^\mu\chibar\big) v_\mu 
   + 8 i g \phi^\dag \lambda 
   + 4 i g \phi^\dag v_\nu \big(\sigma^\mu\sibar^\nu\psi_\mu\big)\Big] \nn \\
&& - \Theta \!\cdot\! \Theta \Big[
  8 i g v_\mu \Dhat^\mu\phi^\dag
    \!-\! 4 \sqrt{2} i g \chibar\!\cdot\!\lambar
    \!+\! 4 i g \phi^\dag \Dhat_\mu v^\mu
    \!+\! \frac{8}{3} g \phi^\dag v_\mu b^\mu
    \!-\! 4 g D \phi^\dag 
    \!+\! 4 g^2 \phi^\dag v_\mu v^\mu\  
  \Big]\nn \ .
\nn\eeqa

The superfield strength tensor associated to a vector superfield reads
\beqa
\label{eq:Wl}
W_\alpha&=&-2g\bigg[i\lambda_\alpha + 
  \Big[
     i (\sigma^{\mu \nu}\Theta)_\alpha (\Fhat_{\mu\nu})
      + \Theta_\alpha D \Big] \\
      &&\hskip .6truecm 
   - \Theta \!\cdot \!\Theta \Big[
        (\sigma^\mu{} \Dhat_\mu \lambar)_\alpha   
      - \frac{i}{2} b_\mu (\sigma^\mu \lambar)_\alpha
      - \frac{i}{2}  M^\ast \lambda_\alpha \Big]
  \bigg] \ ,
\nn\eeqa
and the chiral superfield $\R$ is given by
\beqa
\label{eq:Rl}
\R&=& \R\B + \sqrt 2 \Theta \cdot (\D \R)\B -\frac14 \Theta \!\cdot\! \Theta (\D \cdot\D \R)\B\\
&=& \frac16\Bigg\{-M + \Theta\!\cdot\! \Big[
 -2 (\sigma^{\mu\nu}\psi_{\mu\nu})
    + i \big(\sigma^\mu\psibar_\mu\big) M
    - i \psi_{\mu} b^\mu  \Big] 
 - \Theta\!\cdot\!\Theta \Big[
  \frac12 e_\mu{}^\tm e_\nu{}^\tn R_{\tm \tn}{}^{\mu \nu}\big|
  \nn\\ &&\qquad
  + \frac23 MM^\ast 
  - \frac13 b_\mu b^\mu
  - {i} e_\mu{}^\tm \D_\tm b^\mu
  - \frac12 \psibar_\mu \!\cdot\! \psibar^\mu M 
  - \frac12 \psi_\nu \sigma^\nu \psibar_\mu b^\mu\nn \\
 &&\qquad - {i} \psibar^\mu \sibar^\nu \psi_{\mu \nu}
 - \frac{1}{8}\e^{\mu \nu \rho \sigma} \Big[\psi_\mu \sigma_\sigma
     \psibar_{\nu\rho} + \psibar_\mu \sibar_\sigma \psi_{\nu\rho}\Big]
  \Big] \Bigg\}\nn \ .
\eeqa

\subsection{Super-Weyl rescaling and invariant actions}
We are now ready to give the supergravity action in the superspace language. To construct a supergravity Lagrangian we chose a compact Lie
group $\g$ and a unitary representation ${\frak R}$.  As in section \ref{sec:GAUGE} we introduce the corresponding chiral and vector superfields (see \ref{app:conv} for notations). Next we introduce three
gauge invariant functions: (1) the superpotential $W(\Phi)$ a holomorphic function depending on chiral superfields;   (2) the gauge kinetic function
$h_{ab}(\Phi)$ (where $a,b$ are gauge indices in the adjoint representation of $\g$)
a holomorphic function depending on chiral superfields and (3) the K\"ahler potential  $K(\Phi,\Phi^\dag)$ a real function depending on $\Phi$ and $\Phi^\dag.$ 

Following \eqref{eq:coup} in order to have gauge invariant action we
have to substitute $K(\Phi,\Phi^\dag)$ by $K(\Phi,\Phi^\dag e^{-2gV})$.
Using the previous subsection, and in particular \eqref{eq:EPhi}, the action reads
{\small
\beqa \label{eq:chiralaction} 
  {\cal L} = \int   \d^2 \Theta\  \E \Bigg\{\frac38 (\Dbar\!\cdot\!\Dbar -8 {\cal R})  e^{-\frac13 K(\Phi, \Phi^\dag e^{-2gV})} +
W(\Phi) + 
    \frac{1}{16 g^2}  h_{ab}(\Phi) W^{a \alpha} W_\alpha^b\Bigg\}
    + \text{h.c.} \ .\nn\\[1pt]
\eeqa}

\noi
 Almost all terms of the previous action can be understood directly  from
 the corresponding supersymmetric action passing from flat superspace to
 curved superspace except one term which seems to be surprising at a
 first glance. Indeed, we have to exponentiate the K\"ahler potential.
 This exponentiation is fundamental in order to have  correctly normalised
 kinetic terms as we shall see in the next section.\\
 
 If we consider a  super-Weyl transformation with parameter $\Sigma$,
from [\ref{eq:Sig-RGW}-\ref{eq:SigmaE}] we can deduce the transformation
of the supergravity multiplet that we do not give here
(see \cite{mm}, Chapter 5 for details) and deduce the
transformation of $\E$:
 \be \label{eq:htE}
  \delta_\Sigma \E = 6 \Sigma\E +\partial_{\Theta^\alpha} (S^\alpha \E) \ ,
\ee
where we have introduced the spinorial chiral superfield
\be \label{eq:Salpha}
  S^\alpha= \Theta^\alpha \big[2\Sigma^\dag -\Sigma\big]\B + 
    \Theta\!\cdot\!\Theta \D^\alpha \Sigma\B  \ .
    \ee
    Furthermore if we impose that chiral fields  have a conformal weight $w=0$, \ie,
    $\delta_\Sigma \varphi= w \varphi$ with $w=0$ we have (using \eqref{eq:ht})
 \beqa
 \delta_\Sigma \Phi=-S^\alpha \partial_{\Theta^\alpha} \Phi\ . 
\eeqa
Imposing  that vector superfields have  a vanishing conformal weight too,
after some algebraic manipulations, we obtain for a finite transformation
that the action \eqref{eq:chiralaction} transforms as
\beqa
\label{eq:Wact}
  {\cal L} &\to& \frac38 \int \d^2 \Theta\  \E (\Dbar\!\cdot\!\Dbar -8 {\cal R}) 
    e^{2(\Sigma + \Sigma^\dag)}  \exp\left[-\frac13 K(\Phi, \Phi^\dag e^{ -2gV})\right]  \\
&& + \int  \d^2 \Theta \    {\cal E}e^{6 \Sigma} W(\Phi) + 
    \frac{1}{16 g^2} \int  \d^2 \Theta\  \E h(\Phi)_{ab} W^{a \alpha} W_\alpha^b
    + \text{h.c.} \ .\nn
\eeqa
Thus only the gauge part of the action is conformal invariant.
Finally for further use we give the transformations of physical
fields under a finite Weyl transformations:
\beqa
\label{eq:W-F}
&e_\tm{}^\mu \to e^{(\Sigma + \Sigma^\dag)|} e_\tm{}^\mu\ , \ \ 
e \to  e^{2(\Sigma + \Sigma^\dag)|}  e\ ,  \\
& \chi^i \to e^{(\Sigma-2\Sigma^\dag)|} \chi^i\ , \ \ 
  \lambda_a \to e^{-3 \Sigma|} \lambda_a  \ ,   \ \ 
  \bar \psi_\tm \to e^{(2\Sigma-\Sigma^\dag)|}\Big(\bar \psi_\tm - i
    \sibar_\tm \D\Sigma|\Big) \ ,\nn\\
  &  \Fhat_{\mu \nu} \to e^{-2(\Sigma+\Sigma^\dag)|}\Fhat_{\mu \nu} \ , \ \
   v_\mu^a \to e^{-(\Sigma+\Sigma^\dag)|} v_\mu^a  \ , \ \ 
    D^a \to e^{-2(\Sigma+\Sigma^\dag)|}D^a\nn \ . 
\eeqa

There is another transformation closely related to  super-Weyl transformations and called K\"ahler transformation. A K\"ahler transformation  reads:
\beqa
K(\Phi,\Phi^\dag) &\to& K(\Phi,\Phi^\dag) + F(\Phi) + F^\ast(\Phi^\dag) \ ,\nn\\
W(\Phi) &\to& e^{-F(\Phi)} W (\Phi)\ , \nn
\eeqa
with $F$  gauge invariant and holomorphic.  Considering a K\"ahler transformation with parameter $F= \ln(W)$ we can recast \eqref{eq:chiralaction} on the form
{\small
\beqa
  {\cal L} =\int \d^2 \Theta\  \E \Bigg\{ \frac38 (\Dbar\!\cdot\!\Dbar -8 {\cal R})  e^{-\frac13 {\cal G} (\Phi, \Phi^\dag e^{-2gV})} +1
    + \frac{1}{16 g^2}  h(\Phi)_{ab} W^{a \alpha}
    W_\alpha^b  \Bigg\} + \text{h.c.} \ ,\nn
\eeqa}

\noi
with $ {\cal G} = K + \ln|W|^2$ the generalised K\"ahler potential.
\section{Supergravity action}\label{sec:act}
In this section we give the main steps to compute in components the action \eqref{eq:chiralaction}. The final result is obtained in two steps.
We firstly expand in components the action and eliminate the auxiliary fields. It turns out that the various kinetic terms are not correctly normalised. In order to have a correctly normalised action we have to perform a super-Weyl rescaling. This is perhaps the longest  computation 
in the derivation of the final Lagrangian.

\subsection{The pure supergravity action}
The material  introduced so far enables us to obtain easily the pure supergravity action:
{\small 
\beqa 
 {\cal L}_{\text{pure sugra}} &=& -3 \int \d^2 \Theta\ \E \R + \hc \ ,\nn\\
 &= &
  \frac12 e e_\mu{}^\tm e_\nu{}^\tn R_{\tm\tn}{}^{\mu\nu} \B
  +\frac14 e \e^{\mu\nu\rho\sigma} \Big[ \psi_\mu \sigma_\sigma
    \psibar_{\nu\rho} - \psibar_\mu\sibar_\sigma\psi_{\nu\rho}\Big]
    - \frac13 e \big[ M M^\ast + b_\mu b^\mu \big]
     \ ,\nn
\eeqa 
}

\noi
where the first term corresponds to the usual Einstein-Hilbert Lagrangian for gravity, the second term corresponds to the Rarita-Schwinger  Lagrangian 
describing the spin$-3/2$ gravitino, and the last two terms describe the
auxiliary fields $M$ and $b_\mu$.  The action, by construction, is obviously  invariant under the
supergravity transformation \eqref{eq:tSUGRA}.
\subsection{Coupling matter and gauge sector to supergravity}
The first step is to expand the Lagrangian \eqref{eq:chiralaction}. As stated previously it is a lengthy computation to obtain the final Lagrangian.
The Lagrangian has three parts: (1) one involving the K\"ahler potential, corresponding to the coupling of  matter with gauge interactions and
supergravity, (2) one involving the superpotential and (3) one gauge part. We have to Taylor expand all these terms using the rules of Section
\ref{sec:GP}. The first term is the most involved. We just decompose
the superspace kinetic energy $\Omega$:
\beqa
\Omega(\Phi,\Phi^\dag) =-3 e^{ -\frac13 K(\Phi,\Phi^\dag) }=  W_ I(\Phi^\dag ) W^I(\Phi) \ , \nn
\eeqa
where $W^I$ are holomorphic functions  whilst $W_I$ are anti-holomorphic functions   and the index $I$ corresponds to the label that we assign to each term in such expansion. 
As in usual supersymmetry we have
\beqa
\label{eq:wexp}
 W^I(\Phi) = W^I + \sqrt{2} \Theta\!\cdot\! (W^I_i \chi^i) 
   -  \Theta \!\cdot\! \Theta \Big[W^I_i F^i + \frac12 W^I_{i j}
      \chi^i\!\cdot\!\chi^j\Big] \ ,
   \eeqa
   (see \ref{app:conv} for notations)
 and all functions depend on the scalar part $\varphi$ of $\Phi$.
We then substitute, as in Section \ref{sec:GAUGE}, $\Phi^\dag$ by
$\Phi^\dag e^{-2g V}$ and compute  the  product of the superfields
$
\E  (\Dbar \cdot \Dbar -8 \R) \Big(W_ I(\Phi^\dag e^{-2g V})\Big) W^I(\Phi) .$
Section \ref{sec:GP} is used to calculate the components of $(\Dbar \cdot \Dbar -8 \R) W_ I(\Phi^\dag e^{-2gV})$ using \eqref{eq:wexp}. Since all superfields have a large number of
components, this computation is lengthy,  but not too complicated. 
At this stage $\Omega$ play the r\^ole of the K\"ahler potential
associated to a  K\"ahler manifold. At the end of the computation all terms involving $\Omega$ and its derivatives
regroup to the geometrical terms associated to 
$\Omega$ along the lines of \ref{app:conv}. 
The second term $-$ associated to the superpotential $-$ is trivial and the last one $-$  the gauge sector $-$   presents no major difficulties. 

Having expanded all terms it is straightforward to eliminate the 
auxiliary fields $F^i$ (associated to chiral superfields), $D^a$ (associated
to gauge multiplets) and $M, b_\mu$ (associated to the gravity multiplet).
However it turns out that the final Lagrangian is not correctly normalised, especially for all kinetic terms. For instance the 
Einstein-Hilbert action takes the form
\beqa
\label{eq:EH}
{\cal L}_{E.H} =-e \frac 16 \Omega  e_\mu{}^\tm e_\nu{}^\tn R_{\tm\tn}{}^{\mu\nu} \B \ . 
\eeqa

Now  it comes the most tedious part of the computation. We just have to consider appropriate super-Weyl rescaling in order to obtain  a correctly normalised Lagrangian. This is done in two steps. In a first step
a Weyl rescaling of the graviton
\beqa 
\label{eq:tW}
e_\mu{}^\tm \to \exp{(-\lambda)} e_\mu{}^\mu \ , \ \ \text{with} \ \ 
\exp{(2\lambda)} = -\frac 3 \Omega \ , 
\eeqa 
recasts \eqref{eq:EH} into the   correctly normalised Einstein-Hilbert action, plus additional terms. 
This corresponds to a superconformal transformation with superfield 
\beqa 
\label{eq:Sigdel}
\Sigma_{\text{dilat}}= \Sigma_{\text{dilat}}^\dag = \frac12 \lambda \ . 
\eeqa
Using \eqref{eq:W-F} and 
Eqs. [\ref{eq:ht}, \ref{eq:tOm}, \ref{eq:SigmaE}] we deduce the transformation
of all fields and all their covariant derivatives. We then perform the
corresponding substitution into the Lagrangian obtained after expansion. 
The final Lagrangian is not at all correctly normalised, and
in order to diagonalise the gravitino kinetic term the Weyl rescaling \eqref{eq:tW}
has to be followed by the  gravitino shift 
  \beqa
  \label{eq:tSW}
\psi_\tm \to  \psi_\tm + \frac{\sqrt{2} i}{2} 
     \Omega^{-1}\Omega^\is \sigma_\tm\bar\chi_\is \nn \ ,  
     \eeqa
     corresponding to the superconformal transformation with superfield
     \beqa 
     \label{eq:Sigshift}
     \Sigma_{\text{schift}}= \sqrt 2( -\frac12 \Omega^{-1} \Omega_i \chi^i  \cdot \Theta) \ .
     \eeqa
     Again, using \eqref{eq:W-F} and 
Eqs. [\ref{eq:ht}, \ref{eq:tOm}, \ref{eq:SigmaE}] we deduce the transformation
of all fields and all their covariant derivatives and perform the
corresponding substitution into the Lagrangian.
Note however that among the three terms in \eqref{eq:chiralaction}, the 
rescaling and the shift of the gauge part are straightforward since the Lagrangian is conformal invariant (see \eqref{eq:Wact}). Note also that after the dilation and the shift the kinetic energy $\Omega$ is naturally replaced by the K\"ahler
potential $K$. This means that the scalar fields $\varphi$ parametrise 
the K\"ahler manifold with K\"ahler potential $K$ and not $\Omega$ (see \ref{app:conv}).

The final Lagrangian has many terms that we shall certainly not give in
this short review. The interested reader can   refer to \cite{mm},  Table 6.2 and 6.3 or to \cite{wb}, Appendix G (with almost the same notations and
conventions). The kinetic part of the final Lagrangian contains in a natural manner 
the Einstein-Hilbert and Rarita-Schwinger Lagrangians for the graviton and gravitino,
the Yang-Mills part and its  corresponding fermionic counterpart for gauge interactions
and the fermionic part together with it scalar counterpart describing the
coupling of matter with gauge interactions and supergravity. Covariant derivative are naturally covariant with respect to all transformations.
For instance for fermionic fields the covariant derivative reads:

{
\beqa
\cD_\mu \chi^{i\alpha} &=& e_\mu{}^\tm\Big[\partial_\tm\chi^{i\alpha} +\chi^{i\beta} 
\omega_{\tm\beta}{}^\alpha +
ig v_\tm^a (T_a\chi^\alpha)^i + \Gamma_j{}^i{}_k\chi^{\alpha j} \tD_\tm \phi^k \nn\\
&& \hskip .5truecm -\frac14\big(
K_j \tD_\tm \phi^j - K^\js \tD_\tm \phi^\dag_\js\big)\chi^{\alpha i} \Big] \ , \nn
\eeqa
}

\noi
where $\tD_\tm \phi^i = e_\mu{}^\tm\Big[\partial_\tm\phi^i + i g v_\tm^a (T_a\phi )^i\Big]$. The last two terms are associated to the K\"ahler
symmetries of the K\"ahler manifold.
The interacting part contains also some natural terms: a coupling of
the gravitino with the golstino fundamental in  the study of symmetry breaking, mass terms for fermionic fields,  a bunch of four-fermions interacting terms and the scalar potential. The non-gauge part of the
latter takes the following expression:
\beqa 
V= e^{K} \Bigg[ \D_i W K^i{}_\is \Dbar^\is W^\ast - 3 |W|^2\Bigg] \ ,\nn
\eeqa 
where the covariant derivative of the superpotential is given by
\beqa 
\D_i W= W_i + K_i W \ . \nn
\eeqa
This means that in supergravity, differently from in supersymmetry,
the potential is no longer positive.
We encourage the readers to consult the reference \cite{mm} in order to guide and facilitate their own computations.

Since the  computation of the final Lagrangian   is laborious various alternative procedures have
been proposed, \eg \ using  superconformal techniques as in \cite{fp}.
Essentially, a compensating field is introduced in order to have a conformal invariant action.  At the end  the artificial superconformal symmetry  is broken selecting an appropriate form for the compensating field. The form of  the compensating field is directly related to the
dilation \eqref{eq:Sigdel} and the gravitino shift  \eqref{eq:Sigshift} because they can be combined in an appropriate  superconformal transformation.

\section{Conclusion}
The Lagrangian described in Section \ref{sec:act} is the starting point of many studies. Several applications have been considered in\cite{mm} such  as the study of the super-Higgs mechanisms and no-scale supergravity, the classification of the so-called soft-supersymmetric breaking terms, some applications in particle physics and cosmology. In the latter case supergravity  in a Jordan frame has been studied.

\appendix
\section{Conventions}\label{app:conv}
In this appendix we give conventions used throughout this review.

For parametrisation of points in the superspace:
Lorentz indices are taken to be untilde  $M=(\mu,\alpha,\alphadot)$ and Einstein indices are tilde $\tM =(\tm,\ta,\taa)$. Greek letters $\mu,\nu,\cdots$
are devoted to vector indices whilst spinor indices are taken to be
$\alpha, \beta,\cdots$ for left handed spinors and $\alphadot,\betadot,\cdots$
for right-handed spinors.

The metric is taken to be $\eta_{\mu \nu} = \text{diag}(1,-1,-1,-1)$ for vectors and $\e_{12}= \e_{\dot 1 \dot 2}=1$ and $\e^{12}= \e^{\dot 1 \dot 2}=-1$ for spinors. Spinorial indices are raised and lowered as follows
\beqa
\psi_\alpha = \varepsilon_{\alpha\beta}\psi^\beta \ , \ \ 
 \psi^\alpha =\varepsilon^{\alpha\beta}\psi_\beta \ , \ \ 
 \bar\psi_\alphadot =\varepsilon_{\alphadot\betadot} \bar\psi^\betadot \ , \ \
 \bar\psi^\alphadot =\varepsilon^{\alphadot\betadot} \bar\psi_\betadot \ \ , \nn
 \eeqa
 and the scalar products of spinors are given by
 \beqa 
  \psi\!\cdot\!\lambda  =\psi^\alpha \lambda_\alpha \ , \ \  
  \psibar\!\cdot\!\lambar =\psibar_\alphadot \lambar^\alphadot \ . \nn
\eeqa
 
The Levi-Civita tensor normalisation is  $\e_{0123}=1$.

The Pauli matrices are given by   (with the following spin-index structure)
\beqa
  \sigma^\mu{}_{\alpha\alphadot} = \big( 1, \sigma^i \big) \ ,  \ \ 
  \sibar^\mu{}^{\alphadot\alpha} = \big(1, -\sigma^i \big) \ , \nn
\eeqa 
with $\sigma^i$ the usual Pauli matrices and $\sigma^0$ the two-by-two identity matrix.
Finally the Lorentz generators in the spin representations are taken to be
\beqa 
  (\sigma^{\mu \nu})_\alpha{}^\beta =\frac14 \Big( 
     \sigma^\mu{}_{\alpha\alphadot} \sibar^\nu{}^{\alphadot \beta} - 
     \sigma^\nu{}_{\alpha\alphadot} \sibar^\mu{}^{\alphadot \beta} 
  \Big)\ , \ \ 
  (\sibar^{\mu \nu})^\alphadot{}_\betadot = \frac14 \Big( 
     \sibar^\mu{}^{\alphadot\alpha} \sigma^\nu{}_{\alpha\betadot} - 
     \sibar^\nu{}^{\alphadot\alpha} \sigma^\mu{}_{\alpha\betadot} 
  \Big) \ .  \nn
\eeqa 

The Pauli matrices enable conversion from spinor indices to vectors
index and {\it vice versa}
\beqa
v_{\alpha \alphadot} = \sigma^\mu{}_{\alpha \alphadot} v_\mu \ , \ \ 
v_\mu = \frac 12 \sibar^{\mu \alphadot \alpha} v_{ \alpha \alphadot} \ . \nn
\eeqa

Consider a compact real Lie algebra $\g$ of dimension $n$. We take the physicists conventions for unitary representation ${\frak R}$, \ie, the matrices  $T_a,a=1,\cdots, n$ acting on ${\frak R}$ are hermitian and fulfill
\beqa
\big[T_a,T_b\big]= i f_{ab}{}^c T_ c \ , \nn
\eeqa
with real structure constants $f_{ab}{}^c$.

Chiral superfields are denoted $\Phi^i$ and anti-chiral superfields
$\Phi^\dag_\is$. From the K\"ahler potential we deduce the 
geometric quantities of the corresponding K\"ahler manifold
\beqa
\text{K\"ahler metric} &:& K^\is}{_i = \frac{\partial ^2 K}{\partial \varphi^i \partial \varphi^\dag_\is} \ , \ \ (K^{-1})^i{}_\is \equiv K^i{}_\is\nn\\
\text{Christofell symbols}&:& 
\Gamma_i{}^k{}_j =
K^k{}_\is \frac{\partial^3 K}{\partial \varphi^i\partial \varphi^j \partial \varphi^\dag_\is} \ , \ \ 
\Gamma^\is{}_\ks{}^\js =
K^i{}_\ks \frac{\partial^3 K}{\partial \varphi^\dag_\is\partial \varphi^\dag_\js \partial \varphi^i} \ 
\nn\\
\text{Curvature tensor}&:&
R_i{}^\is{}_j{}^\js
=\frac{\partial^4 K}{\partial \varphi^i \partial \varphi^j \partial \varphi^\dag_\is\partial \varphi^\dag_\js}
- K^\ks{}_k \Gamma_i{}^k{}_j \Gamma^\is{}_\ks{}^\js \ , \nn
\eeqa

We denote $X_i = \partial_{\varphi^i} X$ throughout and similarly for higher order derivatives.

\section*{Acknowledgments}
We thank J. Zanelli  for his useful comments.
This work has been partially funded by project AFB170003 and Fondecyt grant 1180368. The Centro de Estudios Cient\'ificos (CECs) is funded by the Chilean Government through the Centers of Excellence Base Financing Program of Conicyt.

\Urlmuskip=0mu plus 1mu\relax
\bibliographystyle{utphys}
\bibliography{ref}

\end{document}